\documentclass[11pt]{article}
\usepackage{amsmath}
\usepackage{amsthm}
\usepackage{amssymb}

\def\half{\frac{1}{2}}

\def\pl{ \: + }
\def\mi{ \: - }

\newfont{\bbbold}{msbm10 scaled \magstep1}

\def\bbX{\mbox{\bbbold X}}

\def\bbX{\mbox{\bbbold X}}

\newfont{\goth}{eufm10 scaled \magstep1}

\def\C{\Gamma}
\def\d{\delta}\def\D{\Delta}
\def\e{\epsilon}
\def\F{\Phi}\def\vf{\varphi}

\def\S{\Sigma}

\def\th{\theta}

\def\be{\begin{equation}}\def\ee{\end{equation}}
\def\bea{\begin{eqnarray}}\def\eea{\end{eqnarray}}
\def\barr{\begin{array}}\def\earr{\end{array}}

\def\o{\omega}\def\O{\Omega}
\def\del{\partial}

\def\xz{\times}

\def\nab{\nabla}

\def\hL{{\widehat{L}}}

\def\pl{{(+)}}\def\mi{{(-)}} 
\let\la=\label

\def\nn{\nonumber}
\def\bd{\begin{document}}
\def\ed{\end{document}}
\def\ba{\begin{array}}
\def\ea{\end{array}}
\def\bea{\begin{eqnarray}}
\def\eea{\end{eqnarray}}
\def\ft#1#2{{\textstyle{{\scriptstyle #1}\over {\scriptstyle #2}}}}
\def\fft#1#2{{#1 \over #2}}
\newcommand{\eq}[1]{(\ref{#1})}
\newcommand{\w}[1]{\\[0.#1cm]}
\def\eqs#1#2{(\ref{#1}-\ref{#2})}
\def\det{{\rm det\,}}
\def\tr{{\rm tr}}
\newcommand{\hoch}[1]{$\, ^{#1}$}
\newcommand{\tamphys}{\it\small Center for Theoretical Physics,
Texas A\&M University, College Station, TX 77843, USA}
\newcommand{\kings}
{\it\small Department of Mathematics, King's College, London, UK}
\newcommand{\uu}
{\it\small Department of Theoretical Physics, Uppsala, Sweden}
\newcommand{\hip}
{\it\small HIP-Helsinki Institute of Physics, P.O. Box 64 FIN-00014 University of Helsinki,
Suomi-Finland}
\newcommand{\stock}
{\it\small Department of Theoretical Physics, Stockholm, Sweden} \makeatletter
\renewcommand\theequation{\thesection.\arabic{equation}}
\@addtoreset{equation}{section} \makeatother

\newcommand{\auth}
{\large P.S. Howe and V. Stojevic}

\begin{document}

\hfill{KCL-TH-06-05}

\hfill{hep-th/0606270}

\hfill{\today}

\vspace{30pt}

\begin{center}
{\Large{\bf On the symmetries of special holonomy sigma models}} \vspace{30pt}

\auth

\vspace{30pt}

\kings

\vspace{60pt}

{\bf Abstract}

\end{center}

In addition to superconformal symmetry, $(1,1)$ supersymmetric  two-dimensional sigma models on special holonomy manifolds have extra symmetries that are in one-to-one correspondence with the covariantly constant forms on these manifolds. The superconformal algebras  extended by these symmetries close as W-algebras, i.e. they have field-dependent  structure functions. It is shown that it is not possible to write down  cohomological equations for potential quantum anomalies when the  structure functions are field-dependent.  In order to do this it is necessary to  linearise the algebras by treating composite currents as generators  of additional symmetries. It is shown that all cases can be  linearised in a finite number of steps, except for $G_2$ and $SU(3)$.  Additional problems in the quantisation procedure are briefly discussed.


\pagebreak \tableofcontents \setcounter{page}{1}


\section{Introduction}

There has been a long history of interplay between differential geometry and supersymmetric
non-linear sigma models starting with the observation that $N=2$ supersymmetry in two dimensions
requires the sigma model target space to be a K\"ahler manifold \cite{Zumino:1979et}. It was first
pointed out in \cite{Delius:1989nc} that one could construct conserved currents in $(1,1)$ sigma
models given a covariantly constant form on the target space, and in \cite{Odake:1988bh} it was
shown that the $(1,1)$ model on a Calabi-Yau three-fold has an extended superconformal algebra
involving precisely such a current constructed from the holomorphic three-form. In
\cite{Howe:1991ic} symmetries of this type were studied systematically in the classical sigma model
setting; each manifold on Berger's list of irreducible non-symmetric Riemannian manifolds has one
or more covariantly constant forms which give rise to conserved currents and  the corresponding
Poisson bracket algebras are non-linear, i.e. they are of W-symmetry type. Subsequently the
properties of these algebras were studied more abstractly in a conformal field theory framework
\cite{Shatashvili:1994zw,Figueroa-O'Farrill:1996hm} and more recently in topological models
\cite{deBoer:2005pt}.

In this paper we revisit the symmetries of classical (1,1) supersymmetric non-linear sigma models
with target spaces which admit torsion-free connections with special holonomy groups. The structure
of the classical Poisson bracket algebra of currents associated with the covariantly constant forms
and the supercurrent is investigated and it is shown that, in most cases, it  can be linearised by
the inclusion of a finite number of composite currents. The exceptional cases are $SU(3)$ and
$G_2$, possibly the two cases of most interest in string theory. In these cases derivatives of the
original currents are generated and the presence of these suggests that finite linearisations may
not be possible.

The main motivation for studying these symmetries is as a preparation for trying to gain a better
understanding of them at the quantum level. Such an understanding may be of use in the study of
higher-order corrections in string theory in the sigma model context
\cite{Gross:1986iv,Grisaru:1986px,Candelas:1986tz}, a topic which has recently received renewed
attention from the point of view of spacetime supersymmetry \cite{Lu:2004ng}.

Since the symmetry transformations associated with covariantly constant forms of degree greater or
equal to three are non-linear, even for a flat target space, one might anticipate that BRST
techniques would be necessary in their analysis, and since the algebras only close in a
field-dependent way one would also expect that the BV version might be helpful. The idea would be
to use these techniques in the context of the algebraic renormalisation programme
\cite{Piguet:1995er} in order to study possible anomalies in a cohomological framework
\cite{Howe:1990pz}. However, it turns out that this formalism is not sufficient to deal with the
problems we are mainly interested in, namely the sigma model either by itself or in the presence of
external gauge fields. This motivates the search for linearised extensions of the classical special
holonomy W-algebras.

The symmetry transformations associated with covariantly constant forms have the property that
their parameters are chiral in the sense that they depend on half of the worldsheet superspace
coordinates (see (\ref{2.7})). The BRST transformation of the matter field $X^i$ is\footnote{The
deWitt notation is being used temporarily, with repeated indices implying integration as well as
summation over labels. See for example \cite{Gomis:1994he}.},

\begin{equation}
\label{eq:BRST_matter} s X^i = c^A R_A{}^i(X)
\end{equation}

where $c^A$ are the parameter ghosts.  In order that the BRST operator $s$ is nilpotent on $X^i$ we
would like the transformation of the ghosts to take the standard form

\begin{equation}
\label{eq:BRST_ghost} \it{s} c^A = \half c^C c^B f_{BC}{}^A(X) \ ,
\end{equation}

where $f_{BC}{}^A(X)$ are the field-dependent structure functions defined by $[R_B,R_C]=f_{BC}{}^A
R_A$, the $R_A$s being regarded as vector fields on the space of sigma model fields. However, this
is not consistent because one is transforming chiral objects into non-chiral ones. The structure
functions depend on the currents and are only chiral on-shell. This then prevents one from writing
down a classical master equation in the chiral, or superconformal, theory.

One can circumvent this problem by gauging the algebra in which case the ghosts are no longer taken
to be chiral. In the context of a $d=2$ superconformally invariant action  $S_0(X)$, for a single
chiral sector,\footnote{For both sectors the gauging procedure is  more complicated - see
\cite{Schoutens:1990ja, Hull:1993kf}.} this involves the modification of the original action $S_0$
by

\begin{equation}
\label{eq:gauging} S_0(X) \rightarrow S_0(X) +  h^A j_A(X) \ ,
\end{equation}

where $j_A$ are the conserved currents, and the gauge fields $h^A$  are required to transform as

\begin{equation}
 \label{eq:BRST_gauge}
 s h^A = D_{-} c^A + h^C c^B f_{BC}{}^A(X) \ .
\end{equation}

Two further problems now present themselves. The first is that the algebra of transformations may
fail to close on the gauge fields and the ghosts due to relations between the currents which show
up in the Jacobi identities. This is discussed in more detail in section 4. Even when this is not a
problem the BRST transformations (\ref{eq:BRST_matter}, \ref{eq:BRST_ghost}, \ref{eq:BRST_gauge})
involve the background fields, $c^A$ and $h^A$, transforming into expressions containing quantum
fields. It is possible to construct a solution to the classical master equation, but the naive Ward
identities involve insertions of composite operators \cite{Bastianelli:1991yk}. It is not known how
to analyse anomalies using cohomological techniques in this situation and it is therefore
difficult, if not impossible, to carry through the algebraic renormalisation programme order by
order in perturbation theory. On the other hand, a perturbative evaluation of the explicit
non-local expressions involving composite operators is hopelessly difficult in the context of a
generic special holonomy sigma model.

As far as we have been able to ascertain there seems to be no way out of this apart from
linearisation. For the cases where we can establish finite linearisations at the classical level it
should  be possible to analyse the renormalisation of the symmetries of the effective action itself
in perturbation theory. We discuss this further in the final section. However, we note that in
order to analyse anomalies in the current algebra, or in the operator product expansion, sources
for the currents need to be introduced which implies that the gauging (\ref{eq:gauging}) is
necessary. So even though writing down OPE expressions involving composite currents is commonplace
in the abstract CFT,  for special holonomy sigma models it seems that the OPE can really only be
understood explicitly in pertubative renormalisation theory if the algebra can be linearised.
Moreover, as we shall discuss later, some of the finite linearisations turn out to be unstable in
the quantum theory due to the fact that operators which do not appear in the classical Poisson
bracket algebra can be generated by the OPE.



\section{Review of basics}


The action for a $(1,1)$-supersymmetric  sigma model without boundary is

\begin{equation}
S=\int\, dz\, g_{ij} D_+ X^i D_- X^j\ , \label{2.1}
\end{equation}

where $g_{ij}$ is a Riemannian metric on the $n$-dimensional target space $M$. $X^i, i=1,\ldots n$,
is the sigma model field represented in some local chart for $M$ and $z$ denotes the coordinates of
$(1,1)$ superspace $\S$. We shall use a light-cone basis so that
$z=(x^{++},x^{--},\th^{+},\th^{-})$, with $x^{++}=x^0+x^1, x^{--}=x^0- x^1$. $D_+$ and $D_-$ are
the usual flat superspace covariant derivatives which obey the relations

\begin{equation}
D_+^2 = i\del_{++};\qquad D_-^2 = i\del_{--};\qquad \{D_+,D_-\}= 0 \ . \label{2.3}
\end{equation}

We use the convention that $\del_{++} x^{++}=1$. We shall take the superspace measure to be

\begin{equation}
dz:= d^2 x\,D_+ D_- \label{2.4}
\end{equation}

with the understanding that the superfield obtained after integrating over the odd variables (i.e
after applying $D_+ D_-$ to the integrand) is to be evaluated at $\th=0$.

The action \eq{2.1} is invariant under superconformal transformations which act independently on
the left (+) and right (-) light-cone sectors. In the left sector, a superconformal transformation
takes the form

\begin{equation}
\d X^i= 2a_{--}\del_{++} X^i -i D_+ a_{--} D_+ X^i\ , \label{2.7}
\end{equation}

where the parameter $a_{--}$ is chiral, $D_-a_{--}=0$. The corresponding supercurrent is the
energy-momentum tensor

\begin{equation}
T_{+3}:=g_{ij} \del_{++}X^i D_+X^j\ , \label{2.8}
\end{equation}

The current is conserved in the sense that $D_-T_{+3}=0$ on-shell. Similarly, there is a conserved
energy-momentum tensor $T_{-3}$ in the right sector.

We shall say that the target space has special holonomy if there are one or more covariantly
constant forms which reduce the corresponding holonomy groups from $SO(n)$ to two groups $G$ on
Berger's list. These are: $U(m)$ and $SU(m)$ for $n=2m$; $Sp(k)$ and $Sp(k)\cdot Sp(1)$ for $n=4k$;
$G_2$ and $Spin(7)$.

Let $L$ be a vector-valued $l$-form such that the $l+1$-form obtained by lowering the vector index
(taken to be in the first slot) is covariantly constant; this form will also be denoted $L$. (It
should be clear from the context which is meant). The symmetry transformation associated with $L$
is

\begin{equation}
\d_L X^i=a_L L^i{}_L D_+ X^L \label{2.10}
\end{equation}

where the parameter $a_L$ has Lorentz weight $-l$ and Grassmann parity $(-1)^l$ and the multi-index
$L$ denotes $l$ antisymmetrised indices, $L:=[l_1\ldots l_l]$. We shall use the notation $L_2$ to
denote antisymmetrisation over the $l-1$ indices beginning with $l_2$, and so on. Under a general
variation of the field $X$ the change in the action is

\begin{eqnarray}
\d S&=&\int\,dz\, 2g_{ij}\d X^i \nab_- D_+ X^j \nn \w1 &=&-\int\,dz\, 2g_{ij}\d X^i g_{ij} \nab_+
D_- X^j \ . \label{2.11}
\end{eqnarray}

If we substitute \eq{2.10} into the top line of \eq{2.11} we see that $\d S=0$ provided that the
parameter is chiral, $D_-a_L=0$. The corresponding conserved current will also be denoted by $L$;
it satisfies $D_- L=0$ on-shell and is given by

\begin{equation}
L=\frac{1}{l+1} L_{iL} D_+ X^{iL}\ . \label{2.12}
\end{equation}

In order to evaluate the commutator, of two such transformations one needs some algebraic relations
which can be proved for any special holonomy forms. If we set

\begin{equation}\label{2.12.1}
    (L\cdot M)_{iL_2,jM_2}:=L_{ki L_2} M^k{}_{j M_2 }\ ,
\end{equation}

then one can verify that

\begin{eqnarray}
  (L\cdot M)_{i[L_2,jM_2]} &=& (-1)^{l+1}P_{ijL_2 M_2} +\frac{m}{2}g_{i[j}Q_{L_2
  M_2]}\ ,\nn\w1
  (L\cdot M)_{[jL_2,|i|M_2]} &=& (-1)^{l}P_{ijL_2 M_2} +\frac{l}{2}g_{i[j}Q_{L_2
  M_2]}\ ,\nn\w1
  (L\cdot M)_{i[L_2,|j| M_2]} + (i\leftrightarrow j)&=& g_{ij} Q_{L_2
  M_2}-(l+m-2) g_{(i[l_2} Q_{j)L_3 M_2]}\ .
  \la{3.11}
\end{eqnarray}

The tensors $P$ and $Q$ are totally antisymmetric and covariantly constant; in particular cases
they can vanish. Both of them can be used to define $L$-type symmetry transformations, but in the
commutator of two special holonomy transformations, $[\d_L,\d_M]$, $Q$ is combined with the
energy-momentum tensor. After some algebra one finds that

\begin{equation}
[\d_L,\d_M]X^i =\d_P X^i  + \d_K X^i\ , \label{3.12}
\end{equation}

where each term is now a symmetry by itself. The $P$ transformation, which is of standard $L$-type
has parameter $a_P$ given by

\be
 a_P=(-1)^{l+1} m a_M D a_L -(-1)^m l D a_M a_L \ .
 \la{3.13}
\ee

The $K$ transformation is defined as follows. If we set

\be
 K_{i,K}:=g_{i[k_1} Q_{K_2]}\ ,
 \la{3.18}
\ee

where the multi-index $K$ takes on $l+m-1$ values, then it is not difficult to show (for any
covariantly constant antisymmetric tensor $Q$) that

\be
 \d_K X^i=\frac{l+m-1}{l+m-2}\Big(
 a_K K^{\phantom{j}i}_{j\phantom{i}K_2} \del_{++} X^j D_+ X^{K_2}+\frac{i(-1)^k}{k}
 K^i{}_K \nab_+ (a_K D_+ X^K) \Big)
 \la{3.19}
\ee

is a symmetry of the action \eq{2.1}. In fact, the corresponding conserved quantity is the
composite current $T Q$. For the case in hand the parameter $a_K$ is

\be
 a_K=i(-1)^{l+1}\frac{lm(l+m-2) }{2}a_M a_L\ .
 \la{3.20}
\ee


\section{Poisson bracket algebras}


In this section we re-examine the algebra of symmetry transformations for the torsion-free model.
The idea is to try to linearise the W-type algebraic structure by treating any composite currents
as new independent generators. We shall see that for $SU(m), m\geq 4$, $Sp(k)\cdot Sp(1)$ and for
$Spin(7)$ it turns out to be rather simple to do this by including a small number of extra
generators. Since the $Sp(k)$ case is linear anyway ($N=4$ superconformal symmetry), this only
leaves two cases which cannot be linearised straightforwardly, namely $SU(3)$ and $G_2$. The
problem here is that derivatives of the original currents turn up and this interferes with the
finiteness which is otherwise due to the fact that differential forms only have finite degree.

The subject is best studied using Poisson brackets; these were introduced in \cite{Howe:1991ic}.
These brackets are based on the observation that $(1,1)$ superspace factorises,
$\S=\S^\pl\xz\S^\mi$, so that we can view $z^\mi:=(x^{--},\th^-)\in \S^\mi$ as the super-time,
while the other coordinates $z^\pl:=(x^{++},\th^+)\in\S^\pl$ are spatial coordinates on which the
fields depend. On-shell the currents depend only on the latter as they are conserved in super-time.
In the following discussion the minus coordinates are irrelevant, so that we can drop the pluses
from the formulae without loss of clarity. In this section, therefore, $D$ will denote $D_+$ while
$\del$ denotes $\del_{++}$, with $D^2=i\del$.

The basic Poisson bracket (PB) is

\be
 (DX^i(1),DX^j(2)) = g^{ij} \nab_1 \d_{12}
 \la{4.1}
\ee

where $(1,2)$ refer to two different points in $\S^\pl$, $\nab=\nab_+$ and $\d_{12}$ is the
delta-function in $\S^\pl$ which, as there is only one odd coordinate, is Grassmann odd. As all of
the tensors appearing in the currents are covariantly constant, the covariant derivative in the
basic PB can be replaced by the ordinary derivative, and the tensors can be regarded as constants.
With this being understood one can write \eq{4.1} and its corollaries as

\begin{eqnarray}
  (DX^i(1),DX^j(2)) &=& g^{ij} D_1 \d_{12} \nn \w1
   (\del X^i(1),DX^j(2) &=& g^{ij} \del_1 \d_{12} \nn \w1
  (DX^i(1),\del X^j(2) &=& -g^{ij}\del_1\d_{12} \nn\w1
  (\del X^i(1),\del X^j(2)) &=& ig^{ij}\del_1 D_1\d_{12} \ .
  \la{4.2}
\end{eqnarray}

In the following we shall write $j(a)$ to mean a smeared current. For each current the parameter
$a$ has the opposite Grassmann parity, so we have

\be
 j(a)=\int dz\, j(z) a(z)=\int dz\,  a(z)j(z)\ ,
 \la{4.3}
\ee

where $z$ now denotes $z^\pl$.

For any three currents $A,B,C$ and parameters $f,g$ we have

\begin{eqnarray}
  (A(f),BC(g)) &=& (A(f),B(Cg))+(-1)^{BC}(A(f),C(Bg)) \nn\w1
  (AB(f),C(g)) &=& (B(fA),C(g))+(-1)^{AB}(A(fB),C(g))\ .
  \la{4.4}
\end{eqnarray}

The three basic PBs are the superconformal algebra,

\be
 (T(a),T(b))=T(2(\del a b-a\del b) +iDa Db)\ ,
 \la{4.5}
\ee

the PB of the supercurrent with an $L$-current $L$,

\be
 (T(a),L(b))=L(l\del a b-2a\del b +iDa Db)\ ,
 \la{4.6}
\ee

and the PB of two currents, $L,\ M $,

\be
 (L(a_L),M(a_M))=-P(a_P) - T Q(a_K)\ ,
 \la{4.7}
\ee

where $a_P$ and $a_K$ are defined in \eq{3.13}and \eq{3.20} respectively. The programme now is to
compute the PBs of the composite $T Q$ with all of the other currents including itself. It is
easier to do this explicitly case by case (recall that $dim\, M=n$).

\subsection*{$G=U(m);\ n=2m$}

When $G=U(m)$ there is one extra current associated with the complex structure $J$ and the algebra
is just the $N=2$ superconformal algebra. In the present notation this is, in addition to the $N=1$
PB $(T,T)$,

\bea
 (T(a),J(b))&=& J(\del a b-2a\del b + iDa Db)\ ,\nn\w1
 (J(a),J(b))&=& -iT(ab)\ ,
 \la{4.8}
\eea

where the current is $J=\frac{1}{2}J_{ij} DX^{ij}$. The pair $(J,T)$ together make up the $N=2$
supercurrent which can be viewed as a real $N=2$ superfield. There are further $N=2$ multiplets
given by pairs of the form $(J^p,TJ^{p-1})$. The PB algebra generated by these currents closes,

\begin{eqnarray}
  (J^p(a),J^q(b)) &=& -ipq T J^{p+q-2}(ab)\nn \w1
  (TJ^p(a),J^q(b)) &=& \frac{J^{p+q}}{p+q}(q(2q-1)\del a b-2q(p+1)a\del b +iq Da Db)\nn
  \w1
  (TJ^p(a),T J^q(b)) &=& TJ^{p+q}((2q+2)\del a b-(2p+2) a \del b +
  iDa Db)\ .
  \la{4.8.1}
\end{eqnarray}

\subsection*{$G=Sp(k);\ n=4k$}

When $G=Sp(k)$, so that $M$ is a hyperK\"ahler manifold, we have three complex structures
$\{J_r\},\ r=1,2,3$ giving rise to the $N=4$ superconformal algebra with

\be\la{4.9}
 (J_r(a),J_s(b))=-i\d_{rs} T(ab) +\e_{rst} J_t(Da b + a Db)\ .
\ee

\subsection*{$G=Sp(k)\cdot Sp(1);\ n=4k$}

The holonomy groups $Sp(k)\cdot Sp(1)$, which correspond to quaternionic K\"ahler geometries in
$4k$ dimensions, give rise to W-type algebras which admit finite linearisations. There is a set of
three complex structures $\{J_r\}$ but they are not globally defined on the target space. This
means that one cannot define three additional supercurrents. However, there is a covariantly
constant four-form $\o_L=\o_r\wedge \o_r$, where $\o_r$ is the local two-form corresponding to
$J_r$. This gives rise to an $L$-type symmetry and hence we have an $N=1$ superconformal algebra
extended by this current. The full set of currents is given by $\{L^p,TL^q;\ p=1,\ldots k;\
q=1\ldots k-1\}$.

The PB of two L-currents is

\be
 (L(a),L(b))=-4i TL(ab)\ .
 \la{4.9.1}
\ee

Using this result and the PB of $T$ with $L$ one can verify straightforwardly that

\begin{eqnarray}
  (L^p(a),L^q(b)) &=& -4ipq T L^{p+q-1}(ab)\nn \w1
  (TL^p(a),L^q(b)) &=& \frac{L^{p+q}}{p+q}(q(4q-1)\del a b-2q(2p+1)a\del b +iq Da Db)\nn
  \w1
  (TL(a),T L(b)) &=& TL^{p+q}((4q+2)\del a b-(4p+2 a \del b +
  iDa Db)\ .
  \la{4.10.1}
\end{eqnarray}

\subsection*{$G=SU(m);\ n=2m$}

When the holonomy group is $SU(m),m\geq 3$ the target space is a Calabi-Yau manifold. As well as a
complex structure there is a covariantly constant $(m,0)$-form $\O$. We shall work with $L$ and
$\hat L$ which are respectively the real and imaginary parts of $\O$. So the generating set of
currents is $\{T,J,L,{\hat L}\}$, and we shall set $m=l+1$ to be in line with our previous
conventions. $T$ and $J$ generate an $N=2$ superconformal algebra and the pair $\{L,\hL\}$
transform as a chiral $N=2$ superconformal field,

\begin{eqnarray}
  (J(a),L(b))&=& \hL (lDa b + a Db) \nn\w1
  (J(a),\hL(b))&=& -L (lDa b + a Db)\ .
  \la{4.10.2}
\end{eqnarray}

For $m$ even the PBs for the Ls are

\begin{eqnarray}
  (L(a),L(b)) &=& -i l l! T J^{l-1} (ab) \nn\w1
  (\hL(a),\hL(b)) &=& -i l l! T J^{l-1} (ab) \nn \\
  (\hL(a),L(b))&=& l! J^{l-1}(Da b + a Db)\ ,
  \la{4.10.3}
\end{eqnarray}

while for $m$ odd they are

\begin{eqnarray}
  (L(a),L(b)) &=& l! J^{l-1}(Da b- aDb)\nn\w1
  (\hL(a),\hL(b)) &=& l! J^{l-1}(Da b- aDb)\nn\w1
  (\hL(a),L(b))&=&-i l l!TJ^{l-1}(ab)\ .
  \la{4.10.4}
\end{eqnarray}

In both cases the new currents at this level are $K:=TJ^{m-2}$ and $M:=J^{m-1}$. As noted above
this pair forms an $N=2$ supermultiplet of lowest spin $l$. From \eq{4.8.1} we can see that these
currents have vanishing PBs with themselves unless $m=3$ in which case $(TJ,J^2)\sim J^3$.
Moreover, the commutators of powers of $J$ and their products with $T$ with $L$ and $\hL$ are
mostly zero. One has

\be\la{4.11}
 (J^p(a),L(b))= -2i\d_{p2}TL(ab)\ ,
\ee

which comes about using the fact that
\begin{equation}
\label{4.11.1} JL=0  \ ,
\end{equation}
as $\o$ is a (1,1) form while $\o_L$ is the sum of $(m,0)$ and $(0,m)$ parts, and

\be\la{4.12}
 (TJ^p(a),L(b))=\d_{p1}(\del J L + DT \hL)(lab) -T\hL(a Db)\ .
\ee

In order to show this one has to use the identity

\be\la{4.12.1}
 DJL=iT\hL\ .
\ee

We therefore see that only $TJ$ and $J^2$ have non-trivial PBs with $L$ and $\hL$, and $J^2$ only
produces $J^3$. At the classical level, this pair is only generated for $m=3$ and so we conclude
that $SU(m)$ holonomy algebras have finite linearisations for $m\geq 4$. For $m=3$, however, there
is a new operator which involves a derivative.

\subsection*{$Spin(7)$}

The $Spin(7)$ case is similar to $Sp(k)\cdot Sp(1)$. This is an $N=1$ superconformal algebra
extended by a superfield current $L$ of weight $2$. The invariant form is the self-dual four-form
$\F$. The Poisson bracket of $L$ with itself gives rise to the composite current $K=TL$, and the PB
of this with $L$ gives $L^2$ which is simply the current associated with the volume form. The new
composites are superconformal fields and have vanishing PBs with $L$ and each other. Explicitly we
have

\bea
 (T(a), L(b))&=& L(3\del a b -2a \del b + iDa Db) \nn\w1
 (L(a),L(b))&=& 9i TL(Da b)\nn\w1
 (L(a),T L(b))&=& L^2(3 \del a b\frac{3}{2}a\del b -\frac{i}{2}D aDb) \nn\w1
 (L(a),L^2(b))&=& 0\nn\w1
 (TL(a),L^2(b))&=&0\ .
 \la{4.13}
\eea

This result depends only on the algebraic relations $T^2=L^3=TL^2=0$. In fact, this algebra differs
from the $Sp(2)\cdot Sp(1)$ algebra only in the coefficients.

\subsection*{$G_2$}

The other exceptional special holonomy group is $G_2$ in seven dimensions. This has an invariant
three-form $\vf$ and its dual is an invariant four-form; they can be combined to give the $Spin(7)$
four-form $\F$. However, in this case the PB algebra generated by $T$, $L$ (corresponding to $\vf$)
and $M$ (corresponding to $*\vf$) leads to derivatives of the original currents and we are unable
to conclude that there is a finite linearisation. In more detail, the basic PBs of the weight $3/2$
and $2$ currents $L$ and $M$ are

\bea
 (L(a),L(b)&=& 2M(Da b-a Db) \nn\w1
 (L(a),M(b)&=& -18i TL(ab)\nn\w1
 (M(a),M(b)&=&-24i TM(ab)\ .
\eea

The PBs of $L,M$ with $TL,TM$ then give

\bea
 (L(a),TL(b))&=&  TM(\frac{2}{3}Da b-2a Db) + 2(\del L L+DT M)(ab) \nn\w1
 (L(a),TM(b))&=&  \frac{3}{7}LM(6\del ab-2a\del b+ iDa Db) \nn\w1
 (M(a),TL(b))&=&  0 \nn\w1
 (M(a),TM(b))&=& 0\ .
\eea

To show these it is necessary to make use of the obvious algebraic identities, such as $L^2=0$, as
well as the less obvious ones

\bea
 DLL&=& \frac{4i}{3}TM \nn \w1
 LDM&=&-\frac{4}{7}D(LM);\qquad DL M=\frac{3}{7}D(LM)\nn\w1
 L\del M&=&\frac{4}{7}\del(LM);\qquad \ \ \ \del L M=\frac{3}{7}\del(LM) \label{eq:G2_relations}
\eea

The PBs of the bilnears are

\bea
 (TL(a),TL(b))&=&-T(7\del L L +\frac{8}{3}DT M)(ab)\nn\w1
 (TL,TM)&=&(TM,TM)=(TL,LM)=(TM,LM)=(LM,LM)=0  \ .
\eea

After a little algebra one can show that the derivative operator in $(L,TL)$ can be replaced by $A$
which is defined to be the primary part of $DLL+\frac{2}{3}DTM$, i.e. it transforms as a primary of
weight $\frac{7}{2}$ under superconformal transformations and that the right-hand side of $(TL,TL)$
is proportional to $TA$. To this level, we therefore find that the only non-algebraic operator that
occurs is $A$ together with $TA$. Unfortunately, the algebra does not close on this set and higher
derivative operators are generated. It seems highly unlikely that there is a finite linearisation
in this case, or for CY3 which is similar in structure.


\section{Jacobi identities for W-algebras}


In the BRST/BV language the Jacobi identities (JIs) for the commutator are written as

\begin{equation}
\label{eq:jacobi_brst_problem}
 Z^A R_A{}^i :=  c^D c^C c^B\left(f_{BC}{}^E f_{E D}{}^A-R_B{}^j \del_j f_{CD}{}^A\right) R_A{}^i
 = 0 \ .
\end{equation}

If $Z^A$ does not vanish, the JIs imply relations between the generators. This can also be seen in
the JIs for the Poisson bracket. Given a particular field-dependent algebra one can think of these
as abstract relations which hold independently of a particular representation.

In this section we investigate the reducibility relations that occur for special holonomy
W-algebras.  We will show that when $Z^A$ is not zero for the gauged chiral algebras (see
(\ref{eq:gauging})), it is not possible to solve the classical master equation without introducing
further ghosts.

 As noted in the
introduction, linearisation is necessary to analyse current algebras in an interacting CFT. For
linearised algebras $Z^A$ vanishes, but this new obstruction to be discussed here may be of
interest in special circumstances when the composite operator expressions can be evaluated more
easily, or in the context of W-strings, when all the currents are imposed as constraints and the
$h^A$s are treated as quantum fields.

The part of the master equation linear in the ghost antifields, $c^*_A$, contains the term $c^*_A
Z^A$, so if $Z^A$ does not vanish there is a potential obstruction to solving it. It turns out that
for the gauged chiral systems $Z^A$ is a function of the currents, and terms proportional to $h^*
c^*$ can be added to the BV action so that this part of the master equation is satisfied.
Alternatively,  the field dependent closure functions can be set to zero using appropriate terms
proportional to $X^* h^*$.

The closure of the transformations acting on the gauge fields  (\ref{eq:BRST_gauge}) involves the
JIs, and when $Z^A \neq 0$ the algebra closes only modulo certain symmetries which act only on the
gauge fields and which reflect the relations between the currents \cite{Lu:1994sc,
Thielemans:1995hn}. These will be called null symmetries. For example, for the $SU(3)$ case
 the null (BRST) symmetry, ($c^0$ is a parameter ghost),

\begin{equation}
\label{eq:null_symmetry} s h^J = -D(c^0 L) \ \ \ \ \  s h^{\widehat{L}} =  i c^0 T \ \ \ \ \ \ s
X^i = 0 \ ,
\end{equation}

reflects the relation  (\ref{4.12.1}):

\begin{equation}
\label{eq:example_null_relation} DJ L - iT \widehat{L} \equiv 0 \ .
\end{equation}

The gauged chiral action, (\ref{eq:gauging}), is

\begin{equation}
 S_0 + h^L L+ h^{\hL}\hL +h^T T+  h^J J\ .
\end{equation}

It is clear that (\ref{eq:null_symmetry}) is not the unique symmetry  implied by
(\ref{eq:example_null_relation}). There are many possibilities, all related by transformations
which are graded antisymmetric in the equations of motion of the gauge fields. The null symmetries,
modulo transformations of this type, are graded symmetric. Therefore they cannot be absorbed by
adding terms quadratic in $h^*$ to the BV action, and must be introduced as extra symmetries.

Null symmetries are present in any gauged theory with fermions (due to  relations such as $T^2=0$),
and normally they should be ignored. In conventional gauge theories the properness
condition\footnote{See, for example, section 4.3 in \cite{Gomis:1994he} for the definition of
properness.}  ensures the existence of a propagator. In the context of string theory or W-strings
the gauge fields are non-propagating, and the propagator for the matter fields exists even if the solution is not proper.
Nevertheless, it makes sense to impose the properness condition modulo null symmetries.  The reason
is that, if the null symmetries are incorporated into the theory, we face the problem that they are
infinitely reducible. This fact is easy to demonstrate for null symmetries proportional to the
gauge field equations of motion, but it is also true when they are not of this form
\cite{Thielemans:1995hn, Vandoren:1996ku}. Even more seriously, it is not clear which null
symmetries should be included and which should be ignored. If we include one, one might suppose
that we should include them all, but then the theory becomes difficult to manage.

When $Z^A \neq 0$ closure forces the introduction generators for a subset of the possible null
symmetries. In this case it makes sense to relax the properness condition modulo null symmetries to
include this particular subset, but the infinite reducibility still poses a serious obstruction to
understanding the theory \cite{Thielemans:1995hn}. When gauge fields are treated as background
fields a proper solution is not required, and reducibility ghosts need not be
introduced.\footnote{Because the $h^A$s are non-propagating fields, there is also the option that a
non-proper solution would make some sense even when the $h^A$ are treated as quantum fields. This
has not been investigated in the literature.}   In the examples we encounter it is possible to
close the algebra after introducing a finite number of null symmetries.  However,  even for the
non-proper solutions terms increasingly non-linear in the antifields need to be added to the extended
action to solve the master equation at higher orders, and it is not clear whether a finite number
of terms is sufficient.

Null symmetries arise in this manner for many of the special holonomy W-algebras. In the case of
$SU(m)$, using the expressions (\ref{2.8}) and (\ref{2.12}) for the $T$ and $L$ currents,   one
finds the basic relations (\ref{4.11.1}) and (\ref{4.12.1}). These are implied abstractly by the
JIs only in the case of $SU(3)$.  For $SU(4)$ one obtains many relations which follow from
(\ref{4.11.1}) and (\ref{4.12.1}), but which involve higher powers of currents. For example, the
$(L, (L,L))$ Poisson bracket JI implies:

\begin{align}
& J \partial J L \equiv 0 \  \ , \ \   T D_+J \hL \equiv 0 \ \ , \ \  D_+T J \hL \equiv 0 \ ,
\\ \nonumber & J(T \hL - D_+J L) \equiv 0 \ .
\end{align}

For $SU(m)$, $m \geq 5$, the situation changes. The Poisson bracket JIs now involve high enough
powers of currents so that it becomes possible to absorb the null symmetries by terms quadratic in
$h^*$. For example, the $(L,(L,L))$  Poisson bracket JI in $SU(5)$ implies

\begin{equation}
J^3 \hL \equiv 0 \ .
\end{equation}

The $s h^J$ part of the null symmetry vanishes identically, and therefore the $s h^L$ part can be
absorbed by adding a term proportional to $h^*_J h^*_L$ to the BV action. In these cases a proper
solution (modulo null symmetries) to the master equation can be found.

The $G_2$ case is like $SU(3)$, in that the fundamental relations (\ref{eq:G2_relations}) are
implied by the Poisson bracket JIs, so there is a correlation between this point and the problems
with linearisation. For the rest of the special holonomy cases the Jacobi identities do not imply
any relations between the generators. That is to say, for $Spin(7)$ and $Sp(k)\cdot Sp(1)$ there
are no problems with the JIs,  and one can define the classical W-string BRST operator, but in the
context of the OPE one still has the problem that the background fields transform into the quantum
fields.


\section{Discussion}


We have seen that all the classical special holonomy algebras admit finite linearisations, except
for $G_2$ and $SU(3)$.  Thus in all cases except for these we can set up and analyse potential
anomalies using cohomological methods. The simplest case to consider is the chiral symmetry algebra
for flat target space models. The action is

\be
 S=S_0+ X_i^* s X^i + c^*_A s c^A\ ,
 \la{7.1}
\ee

where the ghosts and their antifields are chiral. The ghost term in the action gets no quantum
corrections and is only introduced to tidy up the algebra. Since the theory defined by $S_0$ is
free it is not affected by anomalies, but there could be anomalies in the antifield sector related
to those of the current algebra.

The next step is to gauge models of this type. The action is

\be
 S=S_0+ h^A j_A + X_i^* s X^i + c^*_A s c^A +h^*_A s h^A\ ,
 \la{7.2}
\ee

where the ghosts are now no longer chiral. Again the last two terms do not receive quantum
corrections. Differentiation of the quantum action twice with respect to the gauge fields gives the
two-point function of the currents. Since the OPE, which is straightforward to compute in the free
theory, introduces the missing $J^p$ and $TJ^p$ currents in $SU(m), m\geq 4$, it follows that this
model will be anomalous in these cases. However, for $Spin(7)$ and $Sp(k)\cdot Sp(1)$, one would
expect that these models would be anomaly-free.

In principle, one could apply the same ideas to the gauged and ungauged interacting models.
However, in practice one has to specify a quantisation procedure. The best is the background field
method which allows one to keep track of the geometrical nature of the theory \cite{Friedan:1980jm}
. This can be accomplished by introducing a family of fields $\bbX(s)$ which interpolate between
the background field $X=\bbX(0)$ and the total field $X_t=\bbX(1)$. The field $\bbX(s)$ can be
taken to satisfy the geodesic equation, and the quantum field $Y$ is taken to be the tangent vector
to the geodesic at $s=0$. The background-quantum split involves a non-linear shift symmetry
\cite{Howe:1986vm} which can be shown to be non-anomalous \cite{Blasi:1988sh}. This symmetry
controls the field dependence of the counterterms and can be understood geometrically in terms of
the first jet bundle of the tangent bundle. Symmetries of the sigma model can give rise to linear
transformations of the quantum field if the symmetry variation commutes with differentiation with
respect to $s$. The condition for this to be the case is that the symmetry variation $\d \bbX(s)$
should satisfy the equation of geodesic deviation. Unfortunately, this is not the case for any of
the special holonomy symmetry transformations. This implies that the quantum field has to transform
non-linearly. In principle, therefore, in order to undertake a complete analysis of the anomalies
of the interacting special holonomy sigma models in the background field method one should analyse
these non-linear symmetries for graphs with both background and quantum external lines since the
latter can contribute as subgraphs in the effective action for background fields.

Although a full analysis would involve the above steps it is nevertheless not unreasonable to look
at the potential anomalies of the background field effective action with no external quantum lines.
If one makes a special holonomy transformation of the background field accompanied by the
appropriate transformation of the quantum field the local action in the path integral is invariant
and the change in the quantum field can be absorbed by a field redefinition in the path integral.
Therefore the effective action, defined by

\be
 e^{i\C[X]}=\int\, DY e^{iS[X_t]}\ ,
 \la{7.3}
\ee

should satisfy a potentially anomalous Ward Identity of the form

\be
 W(a_L) \C= \D(a_L) \cdot\C
 \la{7.4}
\ee

where $a_L$ is the chiral parameter for an $L$-type symmetry, $\D(a_L)$ is the anomaly, and

\be
 W(a_L):=\int a_L \d_L X^i \frac{\d}{\d X^i}\ .
 \la{7.5}
\ee

In this situation one has a Wess-Zumino consistency condition of the form

\be
 W(a_L) \D(a_M)-W(a_M) \D(a_L)=\D((L(a_L),M(a_M)))\ .
 \la{7.6}
\ee

Note that for this approach to be valid the algebra in question must be of the linearised type. It
would be of interest to investigate this consistency condition. One might expect that special
holonomy anomalies would be related to the superconformal anomaly. The above approach can be
extended to the gauged case where one would again expect there to be problems for the $SU(m)$ case.

In conclusion, we have seen that it is extremely difficult to analyse the anomalies of special
holonomy symmetry W-algebras in the BV framework. It is essential to consider the (chirally) gauged
models since the ghosts cannot be consistently taken to be chiral themselves. In these models,
however, even in the most favourable cases, such as $Spin(7)$ or $Sp(k)\cdot Sp(1)$, where there
are no difficulties due to problems with the Jacobi identity, one is still faced with the problem
that the background gauge fields and ghosts transform into the quantum fields. It seems that the
only way to avoid this problem in the W-framework is to quantise the gauge fields. Even here, in
many cases, one is faced with the problem of infinite reducibility.

These problems can all be avoided to some extent if one utilises the fact that most of the
classical special holonomy algebras admit finite linearisations. In the background field method one
can then analyse possible anomalies in the effective action with no external quantum lines using
the naive classical Ward identities and consistency conditions. On the other hand the inclusion of
(background) gauge fields can  cause problems due to the fact that the OPE of two currents
generates operators which are not in the original set. The only models free from this problem are
the $Spin(7)$ and $Sp(k)\cdot Sp(1)$ models.

There are only two models, $SU(3)$ and $G_2$, which do not admit finite linearisations. Technically
this is because of the presence of currents in the algebra which then generate others with more and
more derivatives. An interesting observation is that CY algebras admits closed linear subalgebras
generated by $\{T,J,\O\}$ or $\{T,J,\bar\O\}$ where

\be
 L=\O+\bar\O \qquad \hL=i(\O-\bar\O)\ .
 \la{7.7}
\ee

These can be analysed in all cases including CY3. It is the complex currents $\O$ which are related
to the squares of the spectral flow operator so it may be that commuting the spectral flows in two
directions is the source of the problem. The analogous subalgebra in the $G_2$ case is the
tri-critical Ising model \cite{Shatashvili:1994zw}.


\section*{Acknowledgements}


This work was supported in part by EU grant (Superstring theory) MRTN-2004-512194, PPARC grant
number PPA/G/O/2002/00475 and VR grant 621-2003-3454.


\end{document}